\documentclass[useAMS,usenatbib]{mn2e}
\usepackage{ gensymb }

\usepackage{graphicx}
\usepackage[fleqn]{amsmath}
\usepackage{amssymb}
\usepackage{color}

\newcommand{\ie}{i.e. }
\newcommand{\eg}{e.g. }

\newcommand{\p}{\partial}
\newcommand{\dd}{{\rm d}}
\newcommand{\M}{{\cal M}}
\newcommand{\F}{{\cal F}}
\newcommand{\sh}{{\rm sh}}
\newcommand{\rsh}{{r_{\rm sh}}}

\graphicspath{ {figures/} }

\title[Spin-up by SASI spiral modes]
{New insights on the spin-up of a neutron star during core-collapse}

\author[Kazeroni, Guilet \& Foglizzo]
{R\'emi Kazeroni$^{1}$, J\'er\^ome Guilet$^{2}$ and Thierry Foglizzo$^{1}$\\
$^1$ Laboratoire AIM, CEA/DSM-CNRS-Universit\'e Paris Diderot, IRFU/Service d'Astrophysique, CEA-Saclay F-91191, France\\
$^2$ Max-Planck-Institut fur Astrophysik, Karl-Schwarzschild-Str. 1, D-85748 Garching, Germany 
}

\begin{document}

\maketitle

\label{firstpage}

\begin{abstract}

The spin of a neutron star at birth may be impacted by the asymmetric character of the explosion of its massive progenitor. 
During the first second after bounce, the spiral mode of the Standing Accretion Shock Instability (SASI) is able to redistribute angular momentum and spin-up a neutron star born from a non-rotating progenitor.
Our aim is to assess the robustness of this process.
We perform 2D numerical simulations of a simplified setup in cylindrical geometry to investigate the timescale over which the dynamics is dominated by a spiral or a sloshing mode.
We observe that the spiral mode prevails only if the ratio of the initial shock radius to the neutron star radius exceeds a critical value. 
In that regime, both the degree of asymmetry and the average expansion of the shock induced by the spiral mode increase monotonously with this ratio, exceeding the values obtained when a sloshing mode is artificially imposed.
With a timescale of 2-3 SASI oscillations, the dynamics of SASI takes place fast enough to affect the spin of the neutron star before the explosion. 
The spin periods deduced from the simulations are compared favorably to analytical estimates evaluated from the measured saturation amplitude of the SASI wave. 
Despite the simplicity of our setup, numerical simulations revealed unexpected stochastic variations, including a reversal of the direction of rotation of the shock. 
Our results show that the spin up of neutron stars by SASI spiral modes is a viable mechanism even though it is not systematic. 
\end{abstract}

\begin{keywords}
hydrodynamics -- instabilities -- shock waves -- stars: neutron -- stars: rotation -- supernovae: general
\end{keywords}

\section{Introduction}

Neutron star spin at birth is a key parameter to associate pulsars and their progenitors. It carries information about the massive stars which explode in a core-collapse supernova and give birth to a neutron star. 
Natal spins can be estimated via an extrapolation of the spin-down of observed pulsars and the determination of their current age. 
This indicates that a significant fraction of neutron stars are born with modest rotation periods in a broad range from a few tens to a few hundreds milliseconds \citep{popov12,noutsos13}. 
Population synthesis studies suggest that the initial pulsar spin distribution may be modeled by a Gaussian distribution centered around 300 ms \citep{faucher}.
However, early stellar evolution models found that much faster rotation close to breakup would result from the conservation of angular momentum during core collapse \citep{heger00}. 
This large discrepancy in the distribution of pulsar spins at birth has to be resolved by better stellar evolution models and/or a better description of the role played by core collapse dynamics.

Angular momentum transport processes throughout the stellar evolution play an important role by setting the rotation period of the core prior to collapse. 
Recent asteroseismic observations of red-giant stars \citep{beck12,mosser12} seem to require a more efficient angular momentum transport than usually assumed \citep{deheuvels14,cantiello14}.
Considering the transport of angular momentum by magnetic torques driven by a dynamo mechanism, \citet{heger05} estimated  pulsar spins of $10-15$ ms -- slower than previous estimates but still rather fast compared to observational constraints.
Transport of angular momentum by internal gravity waves (see \eg \citet{talon03,lee14}) is also a possibility to explain the distribution of pulsar spins at birth. 
\citet{fuller15} estimated periods of $20$ ms to $400$ ms due to an influx of internal gravity waves during the latest burning stages.
Angular momentum redistribution by hydrodynamic instabilities during the collapse has not been considered in these previous works.

Detailed numerical simulations of core-collapse supernovae show that massive stars do not explode in spherical symmetry \citep{liebendoerfer01} except for the low-mass end\
of supernova progenitors \citep{kitaura06}. Neutrino-driven convection \citep{herant94,janka96} and the Standing Accretion Shock Instability (SASI)\
\citep{blondin03} are able\
to generate large scale asymmetric motions and may play a crucial role in the success of the explosion (see \eg \citet{foglizzo15} for a recent review).
Such asymmetric motions have been confirmed as the consequence of a linear instability using perturbative methods \citep{foglizzo07}. The resulting asymmetric explosions are supported\
by a series of observational evidences. The typical pulsar velocities of a few hundreds km/s \citep{arzoumanian02} is much higher than its progenitor ones. 
A pulsar kick imparted on the neutron star by a global deformation $l=1$ has the potential to explain this velocity distribution \citep{scheck04,scheck06}. 
Also, the distribution of ${}^{44}\rm{Ti}$ observed in Cassiopea A suggests a large scale asymmetric explosion \citep{grefenstette14}.

Unstable modes of SASI can develop sloshing motions of the shock along a symmetry axis and also spiral motions when the axisymmetric constraint is released.
SASI spiral modes can redistribute angular momentum during the collapse and significantly modify the neutron star spin from what could be inferred by angular\
momentum conservation \citep{blondin07a}. Using a simplified adiabatic model \citet{blondin07a} showed that a spiral mode can dominate the dynamics and is able to spin-up\
a neutron star to periods as short as 50 ms even if the progenitor does not rotate. In this idealized scenario, some angular momentum is expelled in the explosion whereas the opposite angular momentum is accreted onto the surface of the neutron star.

This mechanism relies on the dominant action of a spiral mode.
In the linear regime of SASI a sloshing mode can be decomposed as two counter-rotating spiral modes with similar amplitudes and identical growth rates if the progenitor is non-rotating. 
A robust spiral mode may dominate the dynamics only if the symmetry between these counter-rotating spiral modes breaks non-linearly. 
Spiral modes were obtained by \citet{blondin07b} in 2D, \citet{fernandez10} in 3D using an approximation of neutrino cooling.
This numerical result has been confirmed by an experimental shallow-water analog of SASI \citep{foglizzo12}.
Spiral modes dominate occasionally the dynamics in 3D numerical simulation of the collapse of a 27 $M_{\odot}$ progenitor using various approximations of neutrino transport\
\citep{hanke13,couch14,abdi15}.  Employing an analytical approach \citet{guilet14} estimated the amount of angular momentum deposited  when a single spiral mode dominates the dynamics in a non-rotating progenitor and showed that SASI\
has the potential to explain initial pulsar periods of a few tens to a few hundreds milliseconds. The slow end of this range of periods is compatible with the range\ 100 ms - 8 s obtained by \citet{wongwathanarat13} in their simulations of 15 $M_{\odot}$ and 20 $M_{\odot}$ progenitors.
An efficient spin-up mechanism driven by the SASI would require a long-lasting SASI activity up to the point of explosion. \citet{mueller12} observed such a dynamics\
in their general relativistic neutrino hydrodynamics axisymmetric simulation of a $27\,{\rm M_{\odot}}$ progenitor.

However, the non-linear dynamics of SASI may not always be dominated by a spiral mode. Indeed, no separation of angular momentum was obtained by \citet{iwakami08}\
considering a model with both neutrino heating and cooling. Investigating the flow pattern below the shock wave, \citet{iwakami14} showed that either sloshing or\
spiral modes dominate the dynamics depending on the mass accretion rate and neutrino luminosity considered, illustrating the stochasticity of the angular momentum distribution in hydrodynamics simulations.
Spin-up by the SASI can be effective only if the symmetry breaking leading to a single spiral mode has occurred before the explosion takes place. 
Even in this case, the amount of angular momentum accreted is sensitive to the position of the mass cut radius \citep{rantsiou11}.

We propose to study the timescale and rotational consequences of the symmetry breaking, which determines the respective roles of sloshing and spiral modes.
We perform a set of 2D simulations of a simplified model of an accretion flow restricted to the equatorial plane, using cylindrical geometry. 
The set of parameters considered shed some light on the non-systematic and non-deterministic features of the symmetry breaking.

The paper is organized as follows. The physical and numerical models are described in section \ref{sec:methods}.
Section \ref{sec:results} focuses on the properties of the symmetry breaking and the non-linear evolution of SASI to evaluate their consequences on the distribution of angular momentum.
Our simulations are confronted to the dynamics of less idealized environments in section \ref{sec:discussion} in order to discuss the potential role of SASI on the initial neutron star spin.

\section{Methods}
\label{sec:methods}

\subsection{Physical model}

Our model consists of a standing accretion shock centered around a proto-neutron star in a stationary and non-rotating flow. 
For the sake of simplicity we focus our study on the equatorial plane of the collapsing core, using cylindrical coordinates in a setup similar to \citet{yamasaki08}. 
The main advantage of this model is to allow for non-axisymmetric modes of SASI in 2D. The accreting matter is modeled by a perfect gas with adiabatic index $\gamma = 4/3$. 
Above the shock, the supersonic matter falls inwards radially and reaches the shock radius $r_{\rm sh}$ with an incident Mach number $\mathcal{M}_1=5$. 
Below the shock, the matter accretes subsonically onto the surface of the proto-neutron star which radius is noted $r_*$. 
A cooling function is included to mimic the neutrino emission due to electron capture with the approximation $\mathcal{L}_0 \propto P^{3/2}\rho$ \citep{blondin06}, where $\rho$ and $P$ are respectively the density and the pressure. 
Neutrino heating is neglected in order to suppress buoyancy induced convective motions and concentrate on SASI in its simplest form. The gravity is Newtonian and self-gravity is neglected.

The initial solution is computed by solving the time independent continuity, Euler and entropy equations below and above the shock. 
The two solutions are connected by the Rankine-Hugoniot jump conditions neglecting the dissociation of nuclei at the shock.
The resulting dynamics only depends on the ratio of the initial shock to the proto-neutron star radii.
Typical values of the radii ratio $R\equiv\rsh/r_*$ are $R \approx 2$ \citep{couch14} and $R \approx 4$ \citep{marek09}, depending on the progenitor structure.
\footnote{In these works, the ratio $R$ is estimated by averaging the shock and neutrinosphere radii during the SASI domination phase of the dynamics.}
When converting to physical units, we choose a proto-neutron star with radius $r_* = 50\rm{km}$ and mass $M_*=1.3M_{\odot}$, and a constant mass accretion rate $\dot{M}=0.3M_{\odot}\,\rm{s^{-1}}$ as typical values for the stalled shock phase of a core-collapse supernova during the first second after bounce.

\subsection{Numerical model}
\label{sec:numerics}

To run our two-dimensional time-dependent hydrodynamic simulations we use a version of the RAMSES code (\citealt{teyssier02, fromang06}) adapted to cylindrical coordinates $\left(r,\phi\right)$ for which the grid is uniformly spaced.
RAMSES is a second-order finite volume code, which uses the MUSCL-Hancok scheme. The simulations are performed using the HLLD Riemann solver \citep{miyoshi05} and the monotonized central slope limiter.
We set periodic boundary conditions in the azimuthal direction at $\phi = 0$ and $\phi=2\pi$. 
The radial domain covers the interval $r_* \leq r \leq r_{\rm out}$ with an outer boundary $r_{\rm out}/\rsh\sim 3-6$ so that the shock wave does not reach the edge of the domain.
We impose reflexive inner boundary conditions and free outer boundary conditions as in \citet{blondin06,fernandez09b}.
Resolution effects are minimized by fixing the number of radial cells below the initial shock to 150. The total number of radial cells is then in the range [540,1300] depending on the simulation. 
1000 cells are used in the azimuthal direction, which is significantly larger than what can be afforded by current 3D simulations \citep[e.g. 176 cells in][]{hanke13,melson15}.
High resolution is required to properly resolve the steep gradients of the flow dynamics in the vicinity of the proto-neutron star.\

An entropy cutoff is applied to the cooling function in order to avoid the divergence of the numerical solution in the vicinity of the proto-neutron star \citep{fernandez09b,fernandez15}.\
The cooling function is written

\begin{equation}
\mathcal{L} \equiv \mathcal{L}_0\exp{\left[\left(\frac{-S}{k\,S_{\rm min}}\right)^2\right]},
	\label{eq:cooling}
\end{equation}
where $S\equiv(\gamma-1)^{-1}\ln{\left(P/\rho^{\gamma}\right)}$ defines the entropy, $S_{\rm min}$ its value at $r=r_*$ and the value of $k$ is chosen to introduce only minimal modifications to the stationary state flow.

Once the numerical solution has relaxed on the grid for a few hundreds numerical timesteps, two density perturbations at pressure equilibrium are introduced ahead of the shock to trigger two counter-rotating spiral modes $m=\pm1$ or $m=\pm2$ as described in appendix \ref{sec:perturbations}.
Perturbations are decomposed as Fourier modes in the azimuthal direction according to the general form
\begin{equation}
           \delta A(r,t) \equiv {\rm Re}\left(\sum_m \tilde{c}_m(r)\,e^{-i\left(\omega t-m\phi\right)}\right)
\end{equation}
where $\tilde{c}_m(r,t)$ is the complex amplitude. We define the initial $\epsilon-$asymmetry between spiral modes as:
\begin{equation}
 \epsilon \equiv \frac{|\tilde{c}_m|^2 - |\tilde{c}_{-m}|^2}{|\tilde{c}_m|^2 + |\tilde{c}_{-m}|^2}
	\label{eq:asymmetry}
\end{equation}
with $|\epsilon| \leq 1$. Note that $\epsilon=0$ corresponds to a mirror-symmetric sloshing mode and $\epsilon=\pm 1$ to a single spiral mode.

Our aim is to estimate the timescale for a symmetry breaking, after the phase of linear growth which lasts less than $\sim3$ SASI oscillations ($\lesssim 100\,$ms). Two different methods have been developed for that purpose. 
The first one is based on the time evolution of the angular momentum flux through the inner boundary. 
This flux is very close to zero for a sloshing mode and starts to deviate from zero once one of the spiral modes dominates. 
The second method is based on the angular tracking of the minimum shock radius. This point corresponds to one of the triple points that form in the shock wave.
Its rotation rate evolves rather erratically for a sloshing mode but becomes fairly constant for a spiral mode. The two methods are consistent within a SASI period which is sufficient for our study.

Our code has been carefully tested to check that it does not introduce any artificial source of asymmetry.
If the initial density perturbation is mirror-symmetric, \ie $\epsilon=0$, the mirror-symmetry is conserved and the sloshing mode oscillates along a fixed axis.
Moreover, two simulations with opposite initial perturbations, $\epsilon$ and $-\epsilon$, show two dynamical evolutions that remain mirror-symmetric within machine precision.
Besides, the robustness of the code has been tested by comparing the growth rates and oscillatory frequencies measured in our simulations to those obtained with a perturbative analysis by \citet{yamasaki08}. 
The discrepancies are less than $8\%$ for the growth rates and less than $2\%$\
for the oscillatory frequencies. This is similar to the good agreement obtained by \citet{fernandez09b}.

\section{Results}
\label{sec:results}

\subsection{A critical ratio for the symmetry breaking}

\begin{figure*}
\centering
 \includegraphics[width=\columnwidth]{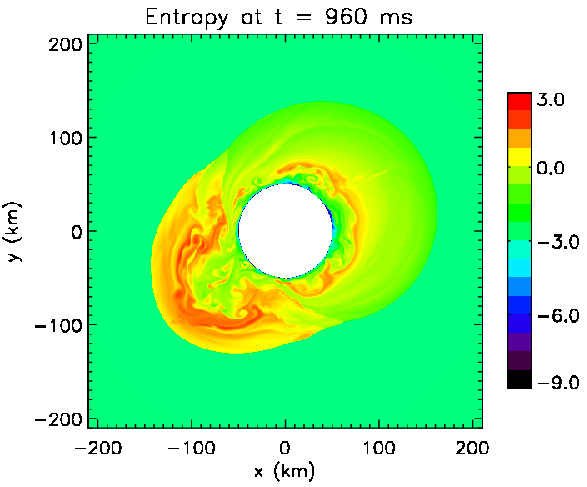}
  \includegraphics[width=\columnwidth]{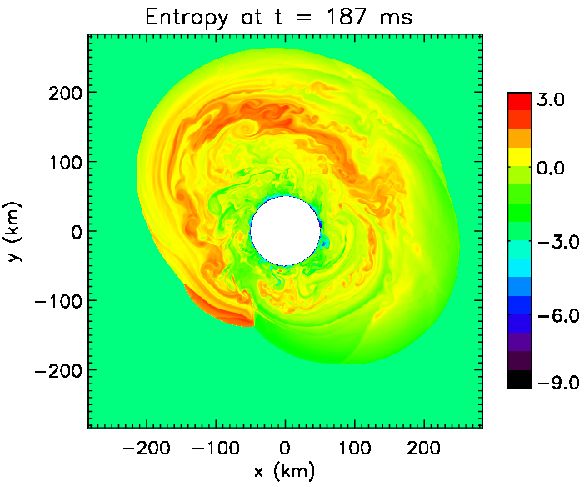}
 \caption{Left: Entropy snapshot at $t=960\,{\rm ms}$ for $R=2$ and $\epsilon=1$. A sloshing motion dominates the non-linear regime despite a spiral perturbation.\
  Right: Entropy snapshot at $t=187\,{\rm ms}$ for $R=3$ and $\epsilon=0.1$. The symmetry breaking has already occurred.\
  (Animated versions of these figures are available in the online journal.)}
             \label{fig:spiral_sloshing}%
\end{figure*}

We performed a total of $80$ simulations varying the two parameters $R$ and $\epsilon$ such that $R=\left\{1.67,\,2,\,2.22,\,2.5,\,3,\,3.5,\,4\right\}$ and $10^{-3}\leq |\epsilon|\leq1$.
Our simulations show that contrary to what was obtained in some studies \citep{blondin07a,fernandez10}, a spiral mode does not always dominate the late evolution.
The ratio $R$ was found to determine whether the symmetry breaking occurs or not.
When $R\leq 2$, the late evolution is dominated by a robust sloshing mode, even if a single spiral mode (\ie $|\epsilon|$=1) was used to perturb the stationary flow (Figure \ref{fig:spiral_sloshing} left).
The azimuthal index of the sloshing mode can either be $m=1$ or $m=2$, depending on the value $R$. In this regime, angular momentum is not significantly redistributed.

A totally different behavior is observed when $R >2$. 
A spiral mode dominates the late evolution (Figure \ref{fig:spiral_sloshing} right) even for weak $\epsilon-$asymmetry, enabling a redistribution of angular momentum. These results raise the question of the mechanism responsible for this symmetry breaking.
The dynamics of SASI observed in our simulations may help to characterize this mechanism as discussed in section \ref{sec:mechanism}.

\subsection{Timescale for the symmetry breaking}\label{sec:symbreak}

\begin{figure}
\centering
 \includegraphics[width=\columnwidth]{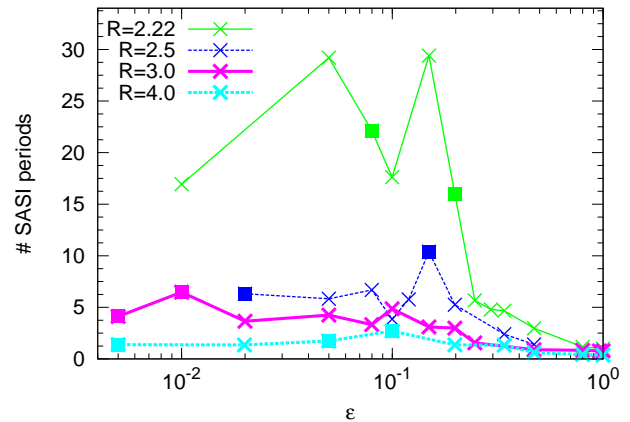}
 \caption{Number of SASI oscillations before reaching a symmetry breaking with respect to the initial asymmetry $\epsilon$. From top to bottom are shown ratios \
  $R=2.22,\,2.5,\,3,\,4$. Square symbols show cases for which the direction of rotation is opposite to the one of the initial asymmetry.}
             \label{fig:timescale}%
\end{figure}

We apply the methods described in section \ref{sec:numerics} to compute the timescale to reach a symmetry breaking as a function of $R>2$ and $\epsilon$ (Figure \ref{fig:timescale}). 
For $R=\left\{2.5,\,3,\,4\right\}$ the symmetry breaking occurs within $2$ to $10$ SASI oscillations in the non-linear phase, which is fast enough to potentially redistribute angular momentum before the explosion. 
In the case $R=2.22$, a spiral mode dominates only after $20$ to $30$ SASI oscillations. This timescale may be too slow to impact the neutron star spin.
The ratio $R=2.22$ illustrates the continuity between a rapid symmetry breaking and an absence of symmetry breaking.

\begin{figure}
\centering
 \includegraphics[width=\columnwidth]{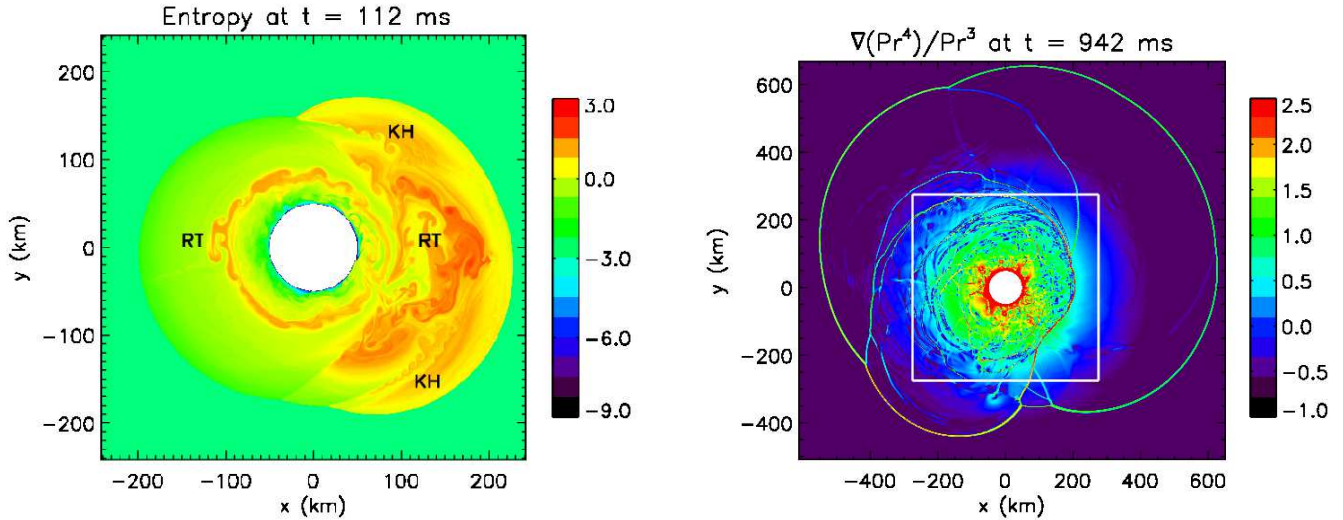}
 \caption{Entropy snapshot at $t=112\,{\rm ms}$ for $R=3$ and $\epsilon=0.01$. Both Kelvin-Helmholtz (\textquotedblleft KH\textquotedblright) and Rayleigh-Taylor (\textquotedblleft RT\textquotedblright) structures are visible after 3 SASI oscillations.\
 These instabilities may add stochasticity before the symmetry breaking between SASI spiral modes.}
             \label{fig:parasites}%
\end{figure}

\begin{figure}
\centering
 \includegraphics[width=\columnwidth]{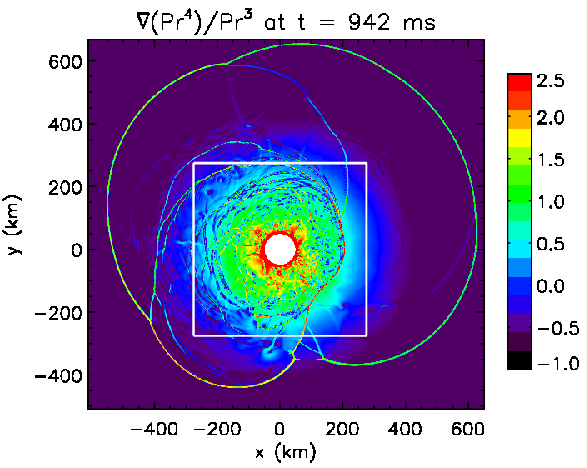}
 \caption{Snapshot of $\left|\nabla\left(P\,r^4\right)\right|/P\,r^3$ at $t=942\,\rm{ms}$ for $R=4$ and $\epsilon=1$. Multiple secondary shocks are present. The solid white lines delimit the domain zoomed in Figure \ref{fig:reverse_mom}.}
             \label{fig:shock}%
\end{figure}

\begin{figure*}
\centering
  \includegraphics[width=\columnwidth]{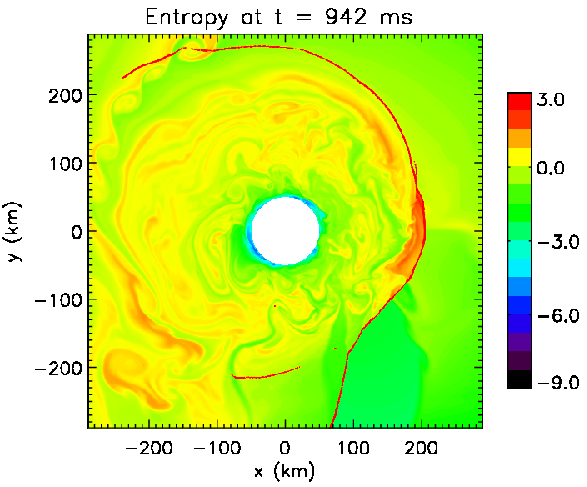}
  \includegraphics[width=\columnwidth]{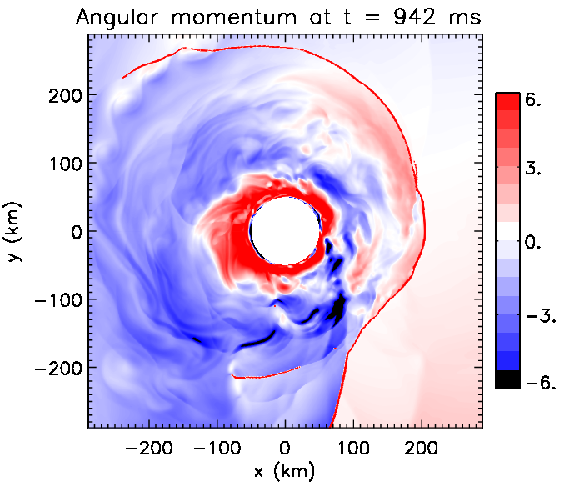}
 \caption{Snapshots of the entropy (left) and the angular momentum (right) at $t=942\,{\rm ms}$ for $R=4$ and $\epsilon=1$ limited to the inner region of the domain\
(see Figure \ref{fig:shock} for a broader view). The red lines display the location of the\
secondary shocks. A secondary shock generates higher entropy material. This region corresponds to positive angular momentum material (red regions below the secondary shock) whereas\
the dynamics is dominated by a spiral mode containing negative angular momentum (blue regions in the angular momentum snapshot).}
             \label{fig:reverse_mom}%
\end{figure*}

The influence of the initial asymmetry on the timescale is not straightforward. If the asymmetry is large enough (\ie $|\epsilon| \geq 0.2$), the timescale decreases with $|\epsilon|$ as would be intuitively expected.
The trend seems rather chaotic and the uncertainties on the timescale are as large as the variability of the results when $|\epsilon| \leq 0.2$.
Furthermore, the direction of rotation of the spiral mode is not always the one determined by the initial asymmetry.
Indeed, approximately half of our simulations with $|\epsilon| \leq 0.2$ show a symmetry breaking in the other direction (square symbols in Figure \ref{fig:timescale}).
The code has been extensively tested to prevent numerical artifacts from inducing asymmetries.
This non-deterministic feature could instead be generated by several non-linear processes which we mention as possible paths towards an explanation. 
The first one is based on the parasitic instabilities, such as Kelvin-Helmholtz and Rayleigh-Taylor that have been proposed to explain the saturation amplitude of the SASI \citep{guilet10}. 
The parasites which develop on SASI spiral modes might modify the asymmetry level in a stochastic way before the symmetry breaking (Figure \ref{fig:parasites}). 
The second non-linear process relies on secondary shocks that arise before the symmetry breaking.
Multiple secondary shocks, shown in Figure \ref{fig:shock}, were witnessed by \citet{fernandez09a}. These shocks may interact with the global advective-acoustic cycle.
The entropy and the vorticity produced by secondary shocks may produce an acoustic feedback in the azimuthal direction opposite to the initial acoustic wave.
This phenomenon might be able to alter the competition between counter-rotating spiral modes and add some stochasticity before a symmetry breaking occurs.
An example is shown in Figure \ref{fig:reverse_mom}, where a secondary shock is able to generate opposite angular momentum well inside the outer shock wave.

\subsection{Reversal of the direction of rotation}
\label{sec:inversion}

\begin{figure}
\centering
 \includegraphics[width=0.8\columnwidth]{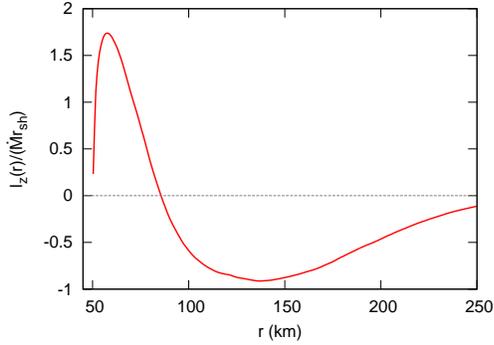}
 \caption{Time averaged angular momentum density profile in the non-linear regime for $R=3$ normalized by $\dot{M}\rsh$. }
             \label{fig:mean_profile}%
\end{figure}

This section and the following one are dedicated to measuring the rotation induced by the spiral mode. We focus on a set of 7 simulations where the radii ratio $R$ is varied and the initial asymmetry is set to $\epsilon=1$. 
The angular momentum density profile
\begin{equation}
 	\label{eq:lz_def}
 l_z(r,t) \equiv r^2\,\int\,\rho v_{\phi}d\phi
\end{equation}
is averaged in time over ten SASI periods in the non-linear phase as shown in Figure \ref{fig:mean_profile}.
Two counter-rotating regions are observed and the radius separating them is labeled $r_0$. If there is no symmetry breaking ($R \leq 2$), the average profile is best defined using the linear phase only.
The angular momentum $L_z(t)$ contained between the radius $r_0$ and the shock wave is computed by:
\begin{equation}
 L_z(t) \equiv \int_{r_0}^{\rsh}\,l_z(r,t)dr.
\end{equation}

\begin{figure}
\centering
 \includegraphics[width=\columnwidth]{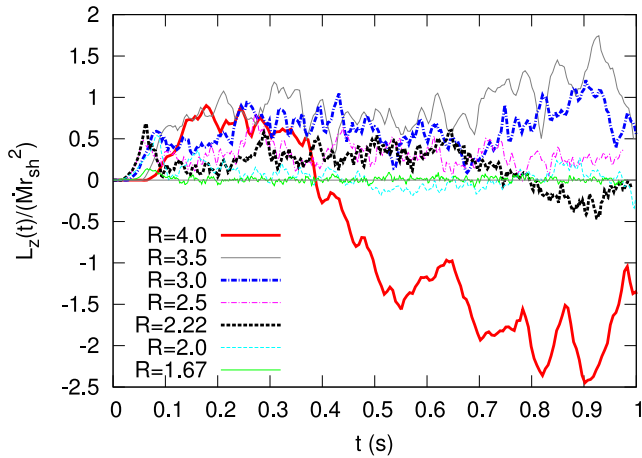}
 \caption{Time evolution of the enclosed angular momentum normalized by $\dot{M}\rsh^2$ for 7 different values of $R$. }
             \label{fig:allLz}%
\end{figure}

Figure \ref{fig:allLz} shows the time evolution of the enclosed angular momentum for our set of simulations.
The cases $R=4$ (solid red line) and $R=2.22$ (dashed black line) exhibit a surprising inversion\
of the direction of rotation of the spiral wave. Both events take place in the fully non-linear regime. 

For $R=2.22$, the change of direction lasts approximately 8 SASI periods during which a sloshing mode dominates.
The angular momentum produced by SASI is very low during that period. Even though a symmetry breaking has occurred, we observe that the non-linear dynamics of the SASI is able to cancel the angular momentum redistribution for a significant time. 
In the case $R=4$, the change of the direction is achieved on a much shorter timescale, less than a SASI period.

The robustness of this behavior was confirmed by repeating these two puzzling simulations varying slightly one parameter such as the numerical resolution or the perturbation amplitude.
This intriguing phenomenon calls for a physical interpretation.
A possibility might be that the secondary shocks discussed in Sect.~\ref{sec:symbreak} break the advective-acoustic cycle and establish temporarily a new one between the second shock and the feedback region.
This process is illustrated by Figure \ref{fig:reverse_mom}. The conditions for this adverse contribution to be able to reverse the direction of rotation remain to be determined. 

\subsection{Estimate of the pulsar spin}
\label{sec:spin}

\citet{guilet14} derived an analytical estimate of the angular momentum redistribution driven by a single spiral mode in spherical geometry. 
This approach has been adapted to the cylindrical geometry in appendix \ref{sec:Lz_cyl}.

\begin{figure}
\centering
 \includegraphics[width=0.8\columnwidth]{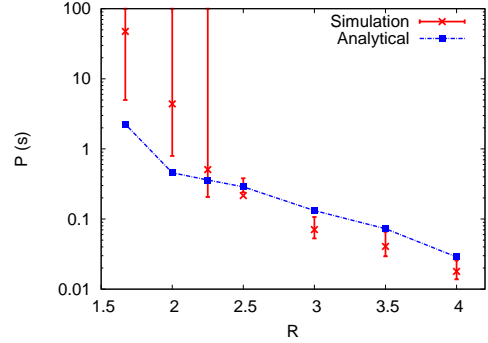}
 \caption{The analytical estimate of the initial neutron star spin period (dot-dashed blue line) is compared to the non-linear regime of our simulations (red bars).\
 The bars refer to the time variation of the amount of angular momentum accreted. For $R<2.5$, various non-linear effects can cancel the angular momentum redistribution and lead to very slowly\
 rotating neutron stars.}
 \label{fig:spin}%
\end{figure} 

Birth periods of neutron stars are inferred from our simulations using a moment of inertia of $I=I_{45} \times 10^{45}{\rm g\,cm^2}$.
Figure \ref{fig:spin} shows a comparison between the analytical estimate (Eq.~\ref{eq:Pns}, using for the spiral mode amplitude the value measured in the simulation as described in next subsection) and the time averaged value in the non-linear regime of our simulations. 
Our results are consistent with the analytical estimates within a factor $2$ for $R \geq 2.5$ and confirm that spiral modes of SASI are able to spin-up a neutron star to periods of tens to hundreds milliseconds. 
The larger discrepancies for $R < 2.5$ are no surprise because the spiral modes do not exist in the non-linear regime ($R = \{ 1.67,\, 2\}$) or a reversal of the direction of rotation takes place during a significant fraction of the non-linear regime ($R=2.22$). 
For that range of parameters, angular momentum redistribution by SASI is inefficient to spin-up the neutron star.

\subsection{Saturation amplitude of SASI}

\begin{figure}
\centering
 \includegraphics[width=\columnwidth]{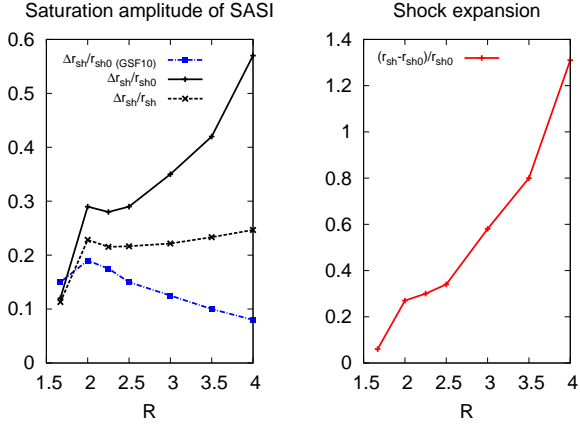}
 \caption{Left: The saturation amplitude of SASI computed by applying the formalism of \citet{guilet10} (dot-dashed blue line) is compared to the one in our simulation for 
 $\epsilon = 1$ and $7$ different values of $R$. In the simulations the saturation amplitude is estimated by averaging the amplitude of the dominant mode over the non-linear regime and normalized by
 the initial shock radius (solid black line) or by the average shock radius in the non-linear regime (dashed black line).
 Right : Variation of the shock radius compared to its initial value $r_{\rm sh0}$ as a function of $R$ for $\epsilon=1$.}
 \label{fig:sat_ampl}%
\end{figure}

The saturation amplitude of SASI $\Delta r$ is a key element of the spin-up by spiral modes because the amount of angular momentum redistributed scales as $\Delta r^2$
(Eq.~\ref{eq:Lz_tot}). The increase of the saturation amplitude with the ratio $R$ (Figure \ref{fig:sat_ampl} left) is consistent with the highest spin obtained for highest values of $R$ (Figure \ref{fig:spin}). 

However, the saturation amplitude obtained by applying the formalism of \citet{guilet10} decreases with increasing $R$ (Figure \ref{fig:sat_ampl} left).
The higher saturation amplitudes observed at large values of $R$ in the simulations indicate that in this regime the parasitic instabilities are not as efficient at stopping the growth of SASI as predicted by \citet{guilet10}.
This suggests either that a more elaborate description of the parasitic instabilities is necessary or that another process is responsible for the saturation of SASI.

The shock expansion due to SASI is increasing with $R$ more steeply than the saturation amplitude: between $R=2$ and $R=4$, it increases by a factor $\simeq4$ while $\Delta r/\rsh$ increases by a factor $\simeq2$ (Figure \ref{fig:sat_ampl}). 
This is consistent with the shock expansion varying quadratically with the saturation amplitude as might be expected for a non-linear effect.
A similar trend was observed by \citet{fernandez09b} with a slighter increase which may be attributed to the geometry difference.
A direct comparison with simulations of steady-state flows including neutrino heating \citep{ohnishi06, iwakami08, iwakami14} is less straightforward because $R$ is also affected by the neutrino luminosity.
Larger ratios may correspond to dynamical evolutions dominated by neutrino driven convection.

\subsection{Differences between spiral and sloshing modes}

\begin{figure}
\centering
 \includegraphics[width=\columnwidth]{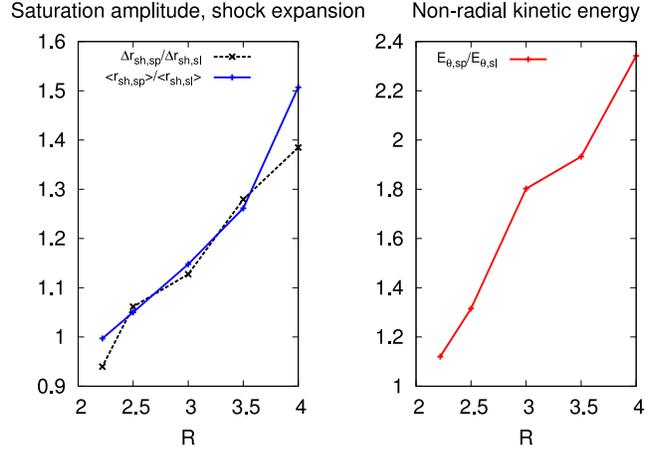}
 \caption{Left: Ratio of the saturation amplitude (dashed black line) and the mean shock radius (solid blue line) between a spiral mode and a sloshing mode for $5$ different
 values of $R$.
 Right : Ratio of the average non-radial kinetic energy in the non-linear regime between a spiral mode and a sloshing mode.
 Note that spiral modes are systematic outcomes for the range of $R$ considered in this figure, unless a mirror-symmetric sloshing is enforced by choosing $\epsilon=0$.
 }
 \label{fig:spi_slo}%
\end{figure}

For $R>2$ the saturation properties of the spiral mode ($\epsilon=1$) are compared with the ones of a sloshing mode obtained by imposing mirror-symmetric initial perturbations ($\epsilon=0$) (Figure \ref{fig:spi_slo} left).
If the ratio $R$ is close to the threshold for symmetry breaking, the average shock radius and the saturation amplitude are almost equal between a sloshing mode and a spiral mode. 
When the ratio $R$ is large enough for the domination of a spiral mode, the shock radius and the saturation amplitude are increased by up to about 40\% compared to the mirror symmetric evolution.

These results confirm the work of \citet{fernandez15} which showed that a spiral mode in 3D may lower the critical neutrino luminosity compared to a sloshing mode in an axisymmetric case that generates less non-radial kinetic energy. 
In our simulations with the highest ratios $R$, spiral modes are indeed able to double the total non-radial kinetic energy compared to sloshing modes (Figure \ref{fig:spi_slo} right), as observed in the simulations without neutrino heating of \citet{fernandez15}.
For smaller ratios $R$ on the other hand, the difference of kinetic energy between spiral and sloshing modes is more modest, suggesting the existence of different regimes. This might be the reason why other groups, in contrast to \citet{fernandez15}, had not found a lowered critical luminosity in 3D simulations exhibiting a spiral mode \citep{hanke13}.

\subsection{A possible path towards the symmetry breaking mechanism}
\label{sec:mechanism}

\begin{figure}
\centering
 \includegraphics[width=\columnwidth]{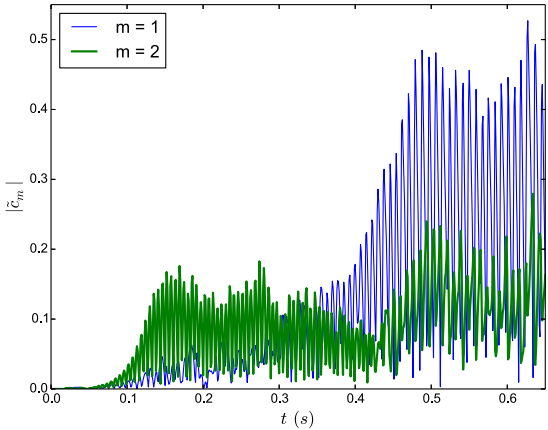}
 \caption{Time evolution of the amplitudes of the Fourier modes $m=1$ (thin blue line) and $m=2$ (thick green line)
 }
 \label{fig:Famp}%
\end{figure}

A description of the physical processes responsible for the symmetry breaking would be helpful to anticipate the efficiency of SASI at spinning-up neutron stars in more realistic models.
A first constraint is that no spiral mode dominates the non-linear dynamics if $R\leq2$.
Additional clues may be inferred from the following properties:

\begin{itemize}
 \item Unlike for $m=1$ modes, the $m=2$ sloshing mode in our setup is never transformed non-linearly into a spiral mode. 
 \item In the case $R=2$, despite a linear domination of the mode $m=2$, the mode $m=1$ eventually prevails due to a non-linear coupling between these modes (Figure \ref{fig:Famp}).
 However, the transition between these modes does not lead to a spiral mode (Figure \ref{fig:spiral_sloshing} left).
 \item Linearly, the mode $m=2$ dominates the mode $m=1$ for $R\leq 2.2$.
 Interestingly, the critical ratio for the symmetry breaking is close to this linear transition.
 \item The efficiency of the symmetry breaking seems to be linked to the difference of saturation amplitudes between the spiral and the sloshing modes.
 A fast symmetry breaking corresponds to a significantly larger amplitude of the spiral mode, while the amplitudes are approximately equal when the symmetry breaking is slow ($R=2.22$, see Figures~\ref{fig:timescale} and \ref{fig:spi_slo} left).
 \end{itemize}

\section{Discussion}
\label{sec:discussion}

\begin{figure}
\centering
 \includegraphics[width=\columnwidth]{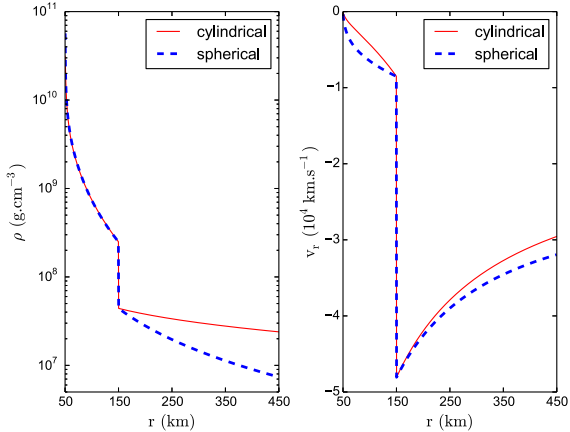}
 \caption{Density profile (left panel) and radial velocity profile (right panel) computed in cylindrical coordinates (solid red line) and in spherical coordinates (dashed blue line). }
 \label{fig:tadv}%
\end{figure}

Several simplifications have been made in our model to study the physics of SASI in its simplest form and less idealized models might modify some aspects of our results.
The dimensionality and the geometry are important points to raise. The density and velocity profiles in cylindrical geometry are compared in Figure \ref{fig:tadv} to those obtained in spherical geometry for the same parameters used in Sect.~\ref{sec:methods}.
In the subsonic region of the flow, the density profile is independent of the geometry but the advection time is shorter in spherical geometry.
Remembering that SASI frequencies and growth rates scale like the advection rate, these quantities are higher in spherical geometry. A symmetry breaking between SASI spiral modes may therefore occur earlier in a 3D spherical model than in 2D cylindrical geometry.
The dimensionality of the model impacts the amount of angular momentum via its dependence on the saturation amplitude. 
The latter was found to be weakly sensitive to the dimensionality \citep{fernandez10,hanke13,fernandez15}. 
However, drawing conclusions on this issue may require to clarify the divergence between the predicted and the measured saturation amplitudes observed in our study (Figure \ref{fig:sat_ampl}).

The initial rotation of the progenitor has been neglected for the sake of simplicity, but could dominate the angular momentum budget if it is fast enough.
Considering the development of spiral modes in a rotating progenitor, \citet{blondin07a} showed that SASI could surprisingly decelerate a neutron star which accretes SASI induced angular momentum opposed to the initial rotation of the stellar core. 
They also showed that this mechanism could even lead to the formation of a counter rotating neutron star.
\citet{yamasaki08} confirmed that rotation favors prograde spiral modes and showed that SASI growth rates depend linearly on the angular momentum of the progenitor.
These results raise the issue of the critical rotation rate of the progenitor above which the neutron star spin at birth is mostly determined by the conservation of initial angular momentum.
A crude estimate can be made by evaluating the angular momentum accreted by a neutron star during the collapse of a non rotating core.
However, the effect of rotation on the saturation amplitude of SASI is still poorly known. The amount of angular momentum redistributed by a spiral mode may increase even if the centrifugal force is negligible.
The mutual influence of the initial rotation and the SASI induced dynamics on the birth period of neutron stars will be addressed in a forthcoming paper.

Taking into account neutrino heating would add a source of stochasticity through the development of neutrino-driven convection, and help address the diversity of explosion paths \citep{fernandez14,cardall15}.
Pre-collapse convective asymmetries may also add stochasticity to post-bounce dynamics and affect the fate of the massive star \citep{couch13,mueller15}.

\citet{iwakami14} explored the diversity of flow patterns behind the stalled shock and observed that the symmetry breaking is not systematic in their SASI-dominated model A with $\dot{M}=0.2\,{\rm M_{\odot}s^{-1}}$, $L_{\nu}=2 \times 10^{52}{\rm erg\,s^{-1}}$ (see table 1 in \citet{iwakami14}).
Changing slightly the numerical resolution or the noise in the initial conditions either lead to a quasi-stationary sloshing or spiral mode. However, their other SASI dominated models exhibit only spiral modes. 
This opens the question of the existence and the value of a critical ratio $R$ for the symmetry breaking in more complex models which may be more subject to stochasticity.
Without neutrino heating in 3D simulations \citet{fernandez10} observed that the amount of angular momentum redistributed by the dynamics of SASI is greatly reduced for $R=1.67$ compared to $R=2$.
These results are consistent with the fact that $R=1.8$ is the transition between $l=1$ and $l=2$ linear modes \citep{foglizzo07}. A similar transition takes place in cylindrical geometry for $R\approx 2.2$.

\section{Conclusion}
\label{sec:conclusion}

A simplified setup in cylindrical geometry has been used to investigate the flow pattern in the non-linear regime of SASI for a non-rotating progenitor. 
A symmetry breaking between counter rotating spiral modes occurs only if the ratio of the initial shock to neutron star radii $R>2$. 
If this condition is satisfied, the dynamics is dominated by a spiral mode, independently of the initial conditions and SASI has the potential to spin-up a neutron star to initial periods of a few tens to a few hundreds milliseconds. 
However, if $R\leq 2$, there is no sign of symmetry breaking and a sloshing mode dominates the dynamics. This case leads to very slowly rotating neutron stars (Figure \ref{fig:spin}). 
These properties set strong constraints on the still unknown mechanism responsible for the non linear symmetry breaking.

The timescale for symmetry breaking of the order of 2-3 SASI oscillations is short enough to affect the angular momentum of the neutron star before the explosion.
This timescale shows stochastic variations when the initial asymmetry is weak (Figure \ref{fig:timescale}).
Memory of the initial perturbations is lost before a symmetry breaking can occur. Moreover the non-linear dynamics of the SASI can lead to a change of the direction of rotation (Figure \ref{fig:allLz}). 
This unexpected result reveals how complex the dynamics can be even in a simplified setup. 
Additional sources of stochasticity are expected from the development of neutrino-driven instabilities as well as pre-collapse convective asymmetries.

Spin-up by SASI may lead to birth periods of neutron stars compatible with observations as proposed by \citep{blondin07a, fernandez10, guilet14}.
The neutron star periods obtained in our simulations are consistent with analytical estimates \citep{guilet14} regardless of the dimensionality of the setup considered.
Diverging conclusions regarding the efficiency of the spin-up mechanism (\eg \citet{iwakami08, rantsiou11}) may be explained by the choice of progenitors or parameters favoring
a dynamics dominated by neutrino-driven convection instead of SASI.

If the shock radius is large compared to the neutron star radius, the saturation amplitude of the spiral mode is larger than the sloshing mode resulting from a mirror-symmetric evolution. 
On the one hand, the larger kinetic energy and shock expansion induced by spiral modes would presumably support a lowered critical neutrino luminosity in 3D, in agreement with \citet{fernandez15}.
On the other hand, when the shock is closer to the neutron star, the saturation amplitude of the spiral mode becomes closer to that of the sloshing mode (Figure~\ref{fig:spi_slo}). 
In that regime, the axisymmetric and 3D dynamics may be expected to be more similar.

The highest saturation amplitudes observed for a large shock radius seem hardly explained by the formalism of \citet{guilet10} and require further investigations.

Initial rotation in the stellar core has been neglected in this study in order to focus on the rotation induced by the spiral mode of SASI.
Rotation is more likely to trigger spiral modes \citep{blondin07a, yamasaki08} which may spin-down the neutron star or even give birth to a counter rotating one.
Nevertheless, the impact of the rotation on the saturation amplitude of SASI and its non-linear dynamics is still poorly known. 
Such a study would help characterize how SASI can affect the mapping between the angular momentum profile of massive stars and the distribution of the initial pulsar spins.
The influence of initial rotation in the core will be addressed in a forthcoming paper in order to disentangle the respective contributions of the initial angular momentum and the dynamical effects of SASI on the pulsar spin at birth.

\section*{Acknowledgements}
The authors are thankful to Marc Joos and Matthias Gonz\'alez for their help with the code. 
We also thank the referee for helping us improve the manuscript.
Numerical simulations were performed using HPC resources from GENCI-TGCC (Grant t2014047094) made by GENCI.
This work is part of ANR funded project SN2NS ANR-10-BLAN-0503.
JG acknowledges support from the Max-Planck--Princeton Center for Plasma Physics.
\appendix

\section{Density perturbations}
\label{sec:perturbations}
In the linear regime, the two spiral modes of index $\pm m$ are triggered by over-densities injected at the outer boundary of the domain:
\begin{equation}
 \delta \rho_{\pm}\left(\theta,t\right) \equiv A\left(1\pm\epsilon\right)\cos{\left(\omega_r\,t \mp m\,\theta\right)}
\end{equation}
where $A$ and $\omega_r$ are the amplitude of the perturbations and the oscillatory frequency of the SASI mode, respectively.
In this formulation $-1 \leq \epsilon \leq 1$ and the sign of $\epsilon$ selects the dominant spiral mode in the linear regime.
The overall perturbation are written as:
\begin{equation}
 \delta \rho\left(\theta,t\right) = H(t)\left(\delta \rho_+\left(\theta,t\right) + \delta \rho_-\left(\theta,t\right) \right)
\end{equation}
where $H(t)$ is a function used to smoothen the perturbation such that:
\begin{equation}
 H(t) \equiv 
\begin{cases}
  \exp\left\lbrace-\left\lbrack\frac{t-\left(t_0+\tau_{\rm adv}/4\right)}{\sigma}\right\rbrack^2\right\rbrace \text{ if } t_0\leq t\leq \tau_{\rm adv}/4\\
  1 \text{ if } \tau_{\rm adv}/4 \leq t \leq 5\,\tau_{\rm adv}/4 \\
  \exp\left\lbrace-\left\lbrack\frac{t-\left(t_0+5\,\tau_{\rm adv}/4\right)}{\sigma}\right\rbrack^2\right\rbrace \text{ if } 5\,\tau_{\rm adv}/4\leq t\leq 6\,\tau_{\rm adv}/4\\
  0 \text{ otherwise}
\end{cases}
\end{equation}
where $t_0$ is the time when the perturbations start to be advected through the outer boundary, $\tau_{\rm adv} = 2\,\pi/\omega_r$ is the advection time and $\sigma$ is a coefficient used to vary the amplitude of $H(t)$ from $10^{-16}$ to $1$ over a timescale $\tau_{\rm adv}/4$.

\section{Angular momentum redistribution by SASI spiral modes in cylindrical geometry}
\label{sec:Lz_cyl}

In this appendix, we adapt the formalism developed by \citet{guilet14} for the angular momentum redistribution by a SASI spiral mode in spherical geometry, to the cylindrical setup considered in this paper. 
Because the derivation of the equations follows very similar steps, we do not reproduce all of them here but highlight the differences linked to the change of geometry. 
The end result takes a very similar form to the spherical geometry, differing only in the numerical factor.

As in the rest of the paper, we consider a 2D accretion flow in cylindrical geometry, assumed to be invariant in the vertical direction and described using cylindrical coordinates $\{ r,\phi \}$. 
The surface integrated angular momentum density is defined in equation~(\ref{eq:lz_def}) (where $\rho$ is to be understood as a vertically integrated surface density). 
While in \citet{guilet14} the surface integration was done on spherical shells, it is here performed on cylinders. Angular momentum conservation can then be written as in \citet{guilet14}
\begin{equation}
\p_t l_z + \p_r \F = 0,
	\label{eq:ang_mom_conservation}
\end{equation}
where $\F$ is the angular momentum flux integrated over a cylindrical surface
\begin{equation}
\F(r,t) \equiv  r^2 \int \rho v_r v_\phi \, \dd \phi.
	\label{eq:def_angmom_flux}
\end{equation}

As in \citet{guilet14}, the flow is described as a stationary background with superimposed small amplitude perturbations
\begin{eqnarray}
\rho(r, \phi, t) & = & \rho_0(r) + \delta\rho(r,\phi, t) + \delta^2\rho(r, \phi, t) + ... \\
v_r & = & v_0 + \delta v_r + \delta^2 v_r + ... \\
v_\phi & = & \delta v_\phi + \delta^2 v_\phi + ...
\end{eqnarray}
where $\delta$ and $\delta^2$ denote first- and second order Eulerian perturbations, respectively, 
with $\delta \gg \delta^2$. With this decomposition, the surface integrated angular momentum density and flux read
\begin{eqnarray}
l_z &=& -\frac{\dot{M}r}{2\pi}\int \left\lbrack \frac{\delta \rho}{\rho_0}\frac{\delta v_\phi}{v_0} + \frac{\delta^2v_\phi}{v_0} \right\rbrack \, \dd\phi , \\
\label{eq:ang_mom_flux_lz}
\F &=& l_z v_0 + T_{Rey} 
\end{eqnarray}
where $\dot{M}\equiv -2\pi r\rho_0v_0$ is the stationary mass flux, and
$T_{Rey}$ is the surface-integrated Reynolds stress
\begin{equation}
T_{Rey}(r,t) = -\frac{\dot{M}rv_0}{2\pi}\int \frac{\delta v_r\delta v_\phi}{v_0^2} \, \dd\phi.
	\label{eq:def_Trey}
\end{equation}
First order perturbations are then decomposed into a superposition of spiral modes with sinusoidal angular dependence with Fourier index $m$ (this replaces the spherical harmonics decomposition used in \citet{guilet14}), and the time-dependence of a plane wave with complex frequency $\omega=\omega_r + i\omega_i$, with $\omega_r$ and $\omega_i$ 
the real and imaginary parts, respectively. The  space and time dependence of
an arbitrary first-order perturbation $\delta A$ is therefore
\begin{equation}
\delta A(r, \phi, t) = \sum_{m} Re\left[\delta \tilde{A}_{m}(r) e^{-i\left(\omega t - m\phi\right)}\right]
\end{equation}
where $\delta\tilde{A}_{m}(r) $ is the complex amplitude\footnote{Contrary to the spherical case, where the azimuthal velocity perturbation $\delta v_\phi$ has a different angular dependence than the rest of 
the variables, the cylindrical geometry allows to describe all the variables with the same Fourier decomposition. When comparing to \citet{guilet14}, this difference of decomposition leads to a numerical factor $im$ 
being included into the complex amplitude of the azimuthal velocity. This --together with the different normalization of Fourier modes compared to spherical harmonics-- is the reason why the Reynolds stress expressions
in equations (\ref{eq:Trey2}) and (\ref{eq:Trey0lm}) differ by a factor $im/4\pi$ from equations (17) and (18) of \citet{guilet14}.}. 
The radial structure and eigenfrequencies of these modes can be computed with a linear analysis as in \cite{yamasaki08}. 

Using this linear eigenmodes decomposition, the Reynolds stress can be written
\begin{equation}
T_{Rey}= -\frac{\dot{M}rv_0}{2}\sum_{m}Re\Big\lbrack \frac{\delta\tilde{v}_{\phi,m}}{v_0}\frac{\delta \tilde{v}_{r,m}^*}{v_0} e^{2\omega_{i,m} t}  \Big\rbrack. \label{eq:Trey2}
\end{equation}
Defining $T_{Rey0,m}$ as the Reynolds stress amplitude of a given mode 
with Fourier index $m$ and with the time dependence scaled out,
\begin{equation}
T_{Rey0,m}(r)= -\frac{\dot{M}rv_0}{2} Re\Big\lbrack \frac{\delta\tilde{v}_{\phi,m}}{v_0}\frac{\delta \tilde{v}_{r,m}^*}{v_0} \Big\rbrack \label{eq:Trey0lm},
\end{equation}
we can write equation~(\ref{eq:Trey2}) as
\begin{equation}
T_{Rey}= \sum_{m} T_{Rey0,m}(r) e^{2\omega_{i,m} t} \label{eq:Trey3}.
\end{equation}

Combining equations (\ref{eq:ang_mom_conservation}), (\ref{eq:ang_mom_flux_lz}), 
and (\ref{eq:Trey3}), angular momentum conservation is written
as a partial differential equation for the evolution of $l_z$
\begin{equation}
\p_t l_z + \p_r(l_z v_0) = - \sum_{m} \p_r T_{Rey0,m}e^{2\omega_{i,m} t}.
	\label{eq:lz_PDE}
\end{equation}
The angular momentum density $l_z$ can therefore be 
written as the sum of contributions from different Fourier modes, with each term given by \citep[see][for more details on the derivation]{guilet14}
\begin{equation}
l_{z0,m} = -\frac{T_{Rey0,m}}{v_0} + \frac{e^{-2\omega_{i,m} \tau_{\rm adv}}}{v_0}\int_{\rsh}^r\frac{2\omega_{i,m} e^{2\omega_{i,m} \tau_{\rm adv}}}{v_0}T_{Rey0,m}\, \dd r, 
	\label{eq:lzlm}
\end{equation}
where $\tau_{\rm adv}(r)=\int_{\rsh}^r \dd r/v_0$ is the advection time from the shock radius $\rsh$ to a 
radius $r < \rsh$.

\subsection{Angular momentum density below the shock}
The angular momentum density below the shock due to a spiral SASI mode follows from equations~(\ref{eq:Trey0lm}) and (\ref{eq:lzlm}),
\begin{equation}
l_{z\sh} = -\frac{T_{Rey0}(\rsh)}{v_\sh} = \frac{\dot{M}\rsh}{2} Re\left(\frac{\tilde{\delta v_\phi}\tilde{\delta v_r}^*}{v_0^2} \right)_\sh \label{eq:lzsh0}
\end{equation}
This expression can be evaluated using the boundary conditions of linear eigenmodes at a shock with a constant dissociation energy
\begin{equation}
\frac{l_{z\sh}}{\dot{M}\rsh} = - m\frac{\omega_r\rsh}{2\pi v_\sh}f(\kappa,\M_1)\left(\frac{\Delta r}{\rsh}\right)^{2},
	\label{eq:lzsh}
\end{equation}
 where $v_\sh$ is the radial velocity below the shock, $\Delta r$ is the amplitude of the shock deformation induced by the SASI spiral mode, $\kappa$ is the compression ratio of the shock, $\M_1$ is the upstream Mach number, and $f(\kappa,\M_1)$ is a dimensionless factor
\begin{equation}
f(\kappa,\M_1) \equiv \pi\left(\kappa -1\right)(1-1/\kappa)\frac{1+1/\M_1^2}{\gamma - (\gamma+1)/\kappa + 1/\M_1^2}.
\end{equation}

\subsection{Approximate expression for the angular momentum contained in a spiral wave}
Using the same method as \citet{guilet14} we obtain an approximate expression for the total angular momentum contained in the spiral wave (i.e. between the shock and the radius where the angular momentum changes sign)
\begin{equation}
L_z  \simeq (\rsh-r_*)l_{z\sh} \simeq m f(\kappa,\M_1) \frac{\omega_r(\rsh-r_*)}{2\pi |v_\sh|}\dot{M}\rsh^2 \left(\frac{\Delta r}{\rsh}\right)^{2}.
 	\label{eq:Lz_tot}
\end{equation}
This is the same equation as \citet{guilet14}, but note that the numerical factor $f(\kappa,\M_1)$ differs by a factor $4\pi$. 
This is mostly due to the different normalization of the Fourier modes considered here compared to the spherical harmonics used in \citet{guilet14}. 
Considering the moment of inertia $I$ of the neutron star, this can be translated into a minimum period of uniform rotation 
\begin{equation}
	\label{eq:Pns}
 P \simeq \frac{2\,\pi\,I}{m f(\kappa,\M_1)} \frac{2\pi |v_\sh|}{\omega_r(\rsh-r_*)}\frac{1}{\dot{M}\rsh^2}\left(\frac{\rsh}{\Delta r}\right)^{2}.
\end{equation}

\bibliography{SASI_spinup}

\bsp
\label{lastpage}

\end{document}